\documentclass[12pt]{article}
\usepackage{latexsym,amsmath,amsfonts,amssymb}
\usepackage{graphicx}
\usepackage{indentfirst}
\usepackage{cite}

\newtheorem{theorem}{Theorem}
\newtheorem{lema}{Lemma}
\newtheorem{definition}{Definition}
\newtheorem{remark}{Remark}

\begin{document}

\begin{center}
 {\Large \bf  Group theoretic analysis of a class of boundary value problems for a nonlinear heat equation}

 \medskip

 {\bf Sergii Kovalenko}

{\it Institute of Mathematics, Ukrainian National Academy of Sciences, \\ Tereshchenkivs'ka Street 3, Kyiv 01601, Ukraine}

 \medskip
 E-mail: kovalenko@imath.kiev.ua
\end{center}

\begin{abstract}
A definition of invariance in Lie's sense for a boundary value problem (BVP) with the basic evolution differential equations is
proposed. A problem of group classification at a wide class of BVPs parameterized by arbitrary elements is formulated and an algorithm for its solution is also proposed. The group classification of a class of BVPs for an (1+1)--dimensional nonlinear heat equation arising from mathematical modeling of heat conduction in semi-infinite body is carried out. An example of invariant solution of a BVP from the class under study is presented.
\end{abstract}



\section{\bf Introduction}

The group-theoretic methods are a modern mathematical apparatus widely used to investigation of different mathematical  models
based both on the ordinary and partial differential equations. The sources of the group analysis of differential equations are found in the fundamental papers of Sophus Lie \cite{Lie1,Lie2}. However, only after the classical works of G. Birkhoff  \cite{Birkhoff} and L.V. Ovsiannikov \cite{Ovs1} have been published, the group-theoretical methods would have been widely used in applications. To the present day, the symmetry properties of many well-known equations from mechanics, hydrodynamics, quantum physics etc. have been studied in detail (see the monographs \cite{Ovs2, Fushchich, Ibr1, Ibr2, Ibr3} and the papers sited therein). This notwithstanding, it should be stressed that only a few papers consider differential equations with initial and boundary conditions. To the best of our knowledge, first works in this direction were published only at the end of 60's years of the last century \cite{mor68, mor69, bl-c}. Nevertheless, interest to the group-theoretic analysis of BVPs have been growing only in the 90's of the twentieth century (see the monographs \cite{Ibr1, Ibr2, Ibr3, Rog-Ames, bl-anco02} and the papers sited therein). At present, the generally-accepted view is that a symmetry (in Lie's sense) of a boundary-value problem must satisfy such three criteria \cite{Rog-Ames, bl-anco02, Ibr3, hydon05}:
\begin{itemize}
    \item[(a)] a symmetry of the governing differential equation;
    \item[(b)] a smooth bijective mapping of the domain to itself;
    \item[(c)] a mapping of the set of boundary data to itself.
\end{itemize}
If a differential equation contains as coefficients arbitrary functions (formally speaking, they can be constants) then the  group classification problem springs up. Such kind of problems was formulated and solved for a class of non-linear heat equations in the pioneering work \cite{Ovs1} (see also \cite{Ovs2}). At the present time, there are algorithms for rigorous solving group classification problems (see, e.g., \cite{ch-ser-ras08, vaneeva09} and the papers cited therein), which  were successfully applied to different classes of partial differential equations. Thus, if a governing equation and/or auxiliary (initial or boundary) conditions contain as coefficients arbitrary functions then one should formulate and solve the group classification problem for such class of BVPs.

In this paper we consider from group-theoretic point of view a class of (1+1)--dimensional nonlinear BVPs modeling the processes of heat transfer in the semi-infinity conductive domain occupying the region $x \geq 0$:
\begin{eqnarray}
& & \frac{\partial u}{\partial t} = \frac{\partial}{\partial
x}\left(d(u)\frac{\partial u}{\partial x}\right), \ \ t > 0, \ x > 0, \label{0.1} \\
& & \quad x = 0: d(u)\frac{\partial u}{\partial x} = q(t), \ \ t > 0, \label{0.3} \\
& & \quad x = + \infty: u = u_{\infty}, \ \ t > 0, \label{0.4}
\end{eqnarray}
where $u(t,x)$ is the unknown temperature field, $d(u) \neq \mbox{const}$ is the nonlinear coefficient of thermal conductivity, $q(t)$ is the some known function describing the heat flux of energy absorbing (or radiating) at the surface $x = 0$, $u_{\infty}$ is the some known temperature at infinity. Additionally, we will assume that all functions arising in problem (\ref{0.1})--(\ref{0.4}) are sufficiently smooth.

It should be noted that BVPs with governing heat conduction equation have a wide field of applications in various branches of
science and technology. That is why the problems of this type were a subject of especial attention of researchers throughout the history of development of the heat conduction theory. In the case of linear heat equation (\ref{0.1}), the problems of the type
(\ref{0.1})--(\ref{0.4}) can be easy solved \cite{Car-Jag, Wang, Luikov}. If the equation is nonlinear than an analysis of the problems like (\ref{0.1})--(\ref{0.4}) is very complicated, and each such a problem requires special investigation. Starting from the classical works of Ya.B. Zel'dovich and G.I. Barenblatt (see the monograph \cite{Zeldovich}), the BVPs for nonlinear heat conduction have been intensively studied by different both exact and approximate methods (see \cite{Crank, Ozisik} and the paper sited therein). However, to the beast of our knowledge, there are a few papers dealing with group-theoretic investigation of problems for nonlinear heat equation and its generalizations \cite{King, Bad-Malek1, Cher, Bad-Malek2, ch-kov-09, ch-kov12a}, but class (\ref{0.1})--(\ref{0.4}) has not yet been the subject of study by researchers working in the field of symmetry analysis of differential equations.

The paper is organized as follows. In Section 2, we formulate a definition of invariance in Lie's sense for a wide class of BVPs. The problem of group classification in classes of BVPs parameterized by arbitrary elements is studied and an algorithm for its solution is also formulated. In Section 3, we perform the complete group classification of the class of BVPs (\ref{0.1})--(\ref{0.4}). in Section 4, we carry out the reduction of the BVP in question in the special case when $d(u) = u^{-\frac{3}{2}}$, $q(t) = q_0$, and $u_{\infty} = 0$ to a BVP of lower dimensionality, and an example of its exact solution is constructed. Finally, we present some conclusions in the last section.

\section{\bf Definition of Lie's symmetry of a BVP for a system of evolution partial differential equations}

Let us consider a BVP of $k$th--order ($k \geq 2$) for a system of $n$ evolution equations with $m + 1$ independent $(t, x)$ (hereafter $x = (x_1, \ldots, x_m)$) and $n$ dependent $u = (u_1, \ldots, u_n)$ variables. We assume that the basic equations
\begin{equation}\label{1}
u_t^j=F^j \left(t, x, u, u_x, \ldots , u_{x}^{(k)}\right), \ j = 1, \ldots, n
\end{equation}
\noindent are defined on a domain ${\Omega} \subset \mathbb{R}^{m+1} $ with smooth boundaries. Hereafter the subscripts $t$ and $x$ denote differentiation w.r.t. these variables, the symbol $u_{x}^{(s)}$ denotes the set of all partial derivatives of $u$ w.r.t. $x$ of order $s$. We, also, will use the symbol $u^{(s)}$ to designate the set of all partial derivatives of $u$ w.r.t. all independent variables $(t, x)$ of order $s$.

Along with system (\ref{1}), we also consider two types of boundary conditions that often appear in applications
\begin{equation}\label{2}
s_a(t,x)=0: \ B^{l}_a \left(t,x, u, u_x, \ldots , u_{x}^{(k_{a}^l)}\right) = 0,\ a = 1, \ldots, p, \, l =1,\ldots,n_a,
\end{equation}
and
\begin{equation}\label{3}
\gamma_c(t,x)=\infty: \ \Gamma^{m}_c \left(t,x, u, u_x, \ldots , u_{x}^{(k_{c}^m)}\right) = 0, \ c = 1, \ldots, q, \, m = 1, \ldots, n_c.
\end{equation}
Here $k_{a}^j < k$ and  $k_{c}^m < k$ are the given numbers, $s_a(t,x)$ and $\gamma_c(t,x)$ are some known functions. We assume
that all functions arising in (\ref{1})--(\ref{3}) are sufficiently smooth.

Consider an $N$--parameter (local) Lie group $G_N$ of point transformations of the variables $(t,x,u)$ in the Euclidean space
$\mathbb{R}^{n+2}$ (open subset of $\mathbb{R}^{n+2}$, respectively) given by equations
\begin{equation}\label{4}
t^{\ast} = T(t,x,\varepsilon), \ \ x_i^{\ast} = X_i(t,x,\varepsilon), \ \ u^{\ast}_j = U_j(t,x,u,\varepsilon),
\end{equation}
where $i = 1, \ldots, m$, $j = 1, \ldots, n$; $\varepsilon = (\varepsilon_1, \ldots, \varepsilon_N)$ are the group parameters. According to the general Lie group theory, one may construct the corresponding $N$--dimensional Lie algebra $L_N$ with the basic  generators
\[
X_\alpha = \xi^0_\alpha \frac{\partial}{\partial t}+\xi^1_\alpha \frac{\partial}{\partial x_1} + \ldots + \xi^m_\alpha \frac{\partial}{\partial x_m} + \eta^1_\alpha \frac{\partial}{\partial u^1}+ \ldots +\eta^n_\alpha \frac{\partial}{\partial u^n}, \ \alpha = 1,2, \ldots, N,
\]
where
$\xi^0_\alpha = \left. \frac{\partial T(t,x,\varepsilon)}{\partial \varepsilon_\alpha}\right \vert_{\varepsilon = 0}, \ \xi^i_\alpha = \left. \frac{\partial X_i(t,x,\varepsilon)}{\partial \varepsilon_\alpha}\right \vert_{\varepsilon = 0}, \ \eta^j_\alpha = \left. \frac{\partial U_j(t,x,u,\varepsilon)}{\partial \varepsilon_\alpha}\right \vert_{\varepsilon = 0}$.

\begin{definition}\label{def1}
The boundary value problem of the form (\ref{1})--(\ref{3}) is called to be invariant w.r.t. the Lie group $G_N$ (\ref{4}) if:
\begin{itemize}
\item[1)] the manifold, determined by the system of equations (\ref{1}) in the space of variables $\left(t,x,u, \ldots, u^{(k)}\right)$, is invariant w.r.t. the $k$th--order prolongation of the group~$G_N$;
\item[2)] each  manifold, determined by conditions (\ref{2}) with any fixed number $a$ in the space of variables $\left(t,x,u,\ldots, u^{(k_a)}\right)$, where $k_a = \max \{k_{a}^l, \ l = 1, \ldots, n_a \}$, is invariant w.r.t. the $k_a$th--order prolongation of the group $G_N$ ;
\item[3)] each  manifold, determined by conditions (\ref{4}) with any fixed number $c$ in the space of variables $\left(t,x,u, \ldots, u^{(k_c)}\right)$, where $k_c = \max \{k_{c}^m, \ m = 1, \ldots, n_c \}$, is invariant w.r.t. the $k_c$th--order prolongation of the group $G_N$ .
\end{itemize}
\end{definition}

\begin{definition}\label{def2}
A group of invariance of BVP (\ref{1})--(\ref{3}) is called the maximal group of invariance (MGI) if any group of invariance of the BVP in question is contained in the MGI as a subgroup.
\end{definition}

It is to be noted that the notion of maximal algebra of invariance (MAI) of BVP (\ref{1})--(\ref{3}) can be formulated in a quite similar way. It immediately follows from Definitions \ref{def1} and \ref{def2} the algorithm to seek for the MGI of some BVP of the form (\ref{1})--(\ref{3}):
\begin{itemize}
\item[$I)$] using the classical Lie algorithm, to find the MGI of the system of differential equations~(\ref{1});
\item[$II)$] to find the maximal subgroup of the Lie group obtained in item $I)$ under which the boundary conditions (\ref{2}) and (\ref{3}) are simultaneously invariant.
\end{itemize}

In Section 4 we will use the notion of invariant solution of BVP (\ref{1})--(\ref{3}). So, the relevant definition must be done.

\begin{definition}\label{def3}
$u = \Phi(t,x)$, with components $u_j = \Phi_j(t,x), \,j = 1, \ldots, n$, is an invariant solution of BVP (\ref{1})--(\ref{3}) corresponding to the Lie group (\ref{4}) if the following conditions take place:
\begin{itemize}
\item[(i)] $u = \Phi(t,x)$ satisfies equation (\ref{1}) and the boundary conditions (\ref{2}) and (\ref{3});
\item[(ii)] the manifold ${\cal{M}} = \{u_j = \Phi_j(t,x), \ j = 1, \ldots, n \}$ is an invariant manifold of the Lie group~ (\ref{4}).
\end{itemize}
\end{definition}

Let us now consider a class of BVPs of the form (\ref{1})--(\ref{3}):
\begin{eqnarray}
& & u_t^j = F^j \left(t, x, u, u_x, \ldots , u_{x}^{(k)}, \theta \right), \label{2.11}\\
& & s_a(t,x) = 0: \ B^{l}_a \left(t,x, u, u_x, \ldots , u_{x}^{(k_{a}^l)}, \vartheta^a \right) = 0,\label{2.12}\\
& & \gamma_c(t,x) = \infty: \ \Gamma^{m}_c \left(t,x, u, u_x, \ldots , u_{x}^{(k_{c}^m)}, \zeta^c\right) = 0, \label{2.13}
\end{eqnarray}
where
\[
\theta = \left(\theta_1\left(t, x, u, u_x, \ldots, u^{(k)}_x \right), \ldots, \theta_r\left(t, x, u, u_x, \ldots, u^{(k)}_x \right) \right),
\]
\[
\vartheta^a = \left(\vartheta^a_1\left(t, x, u, u_x, \ldots, u^{(k_a)}_x \right), \ldots, \vartheta^a_{r_a}\left(t, x, u, u_x, \ldots, u^{(k_a)}_x \right) \right),
\]
\[
\zeta^c = \left(\zeta^c_1\left(t, x, u, u_x, \ldots, u^{(k_c)}_x \right), \ldots, \zeta^c_{r_c}\left(t, x, u, u_x, \ldots, u^{(k_c)}_x \right) \right)
\]
are arbitrary elements satisfying several auxiliary equations and (or) inequalities; the other designations are similar those used in BVP (\ref{1})--(\ref{3}).

\begin{definition}\label{def4}
An equivalence transformation of the class of BVPs (\ref{2.11})--(\ref{2.13}) is a non-degenerate local change of variables
\[
\bar{t} = \tau(t, x,u), \ \ \bar{x} = \varphi(t, x, u), \ \ \bar{u} = \psi(t, x, u),
\]
carrying every BVP of the form (\ref{2.11})--(\ref{2.13}) into a BVP of the same form
\begin{eqnarray}
& & \bar{u}_{\bar{t}}^j=F^j \left(\bar{t}, \bar{x}, \bar{u}, \bar{u}_{\bar{x}}, \ldots , \bar{u}_{\bar{x}}^{(k)}, \bar{\theta}\right), \nonumber\\
& & s_a(\bar{t},\bar{x})=0: \ B^{l}_a \left(\bar{t},\bar{x}, \bar{u}, \bar{u}_{\bar{x}}, \ldots ,
\bar{u}_{\bar{x}}^{(k_{a}^l)}, \bar{\vartheta}^a\right) = 0,\nonumber\\
& & \gamma_c(\bar{t},\bar{x})=\infty: \ \Gamma^{m}_c \left(\bar{t},\bar{x}, \bar{u}, \bar{u}_{\bar{x}}, \ldots ,
\bar{u}_{\bar{x}}^{(k_{c}^m)}, \bar{\zeta}^c\right) = 0. \nonumber
\end{eqnarray}
The functions $\bar{\theta}\left(\bar{x}, \bar{u}, \bar{u}_{\bar{x}}, \ldots, \bar{u}^{(k)}_{\bar{x}} \right)$, $\bar{\vartheta}^a\left(\bar{t}, \bar{x}, \bar{u}, \bar{u}_{\bar{x}}, \ldots, \bar{u}^{(k_a)}_{\bar{x}} \right)$ and $\bar{\zeta}^c\left(\bar{t}, \bar{x}, \bar{u}, \bar{u}_{\bar{x}}, \ldots, \bar{u}^{(k_c)}_{\bar{x}} \right)$ representing the new arbitrary elements, may be, in general, different from the original functions $\theta\left(x, u, u_x, \ldots, u^{(k)}_{x} \right)$, $\vartheta^a\left(t, x, u, u_x, \ldots, u^{(k_a)}_{x} \right)$ and $\zeta^c\left(t, x, u, u_x, \ldots, u^{(k_c)}_{x} \right)$, respectively.
\end{definition}

The set of all such equivalence transformations of the class of BVPs (\ref{2.11})--(\ref{2.13}) forms a group called the equivalence group. Hereafter we denote one by $E_{\mathrm{eq}}^{\mathrm{BVP}}$.

Let now the functions $\theta$, $\vartheta^a$ and $\zeta^c$ are fixed; another words, formulae (\ref{2.11})--(\ref{2.13}) prescribe some BVP. Let the MGI of such BVP is denoted by $G^{\mathrm{BVP}}_{\mathrm{max}}\left(F, \theta, \vartheta^a, \zeta^c \right)$, and $A^{\mathrm{BVP}}_{\mathrm{max}}\left(F, \theta, \vartheta^a, \zeta^c \right)$ is the respective MAI.

\begin{definition}\label{def5}
The common part of $G^{\mathrm{BVP}}_{\mathrm{max}}\left(F, \theta, \vartheta^a, \zeta^c \right)$, i.e.
\[
G^{\mathrm{ker}} = \bigcap_{\theta, \vartheta^a, \zeta^c} G^{\mathrm{BVP}}_{\mathrm{max}}\left(F, \theta, \vartheta^a, \zeta^c \right)
\]
where $\theta$, $\vartheta^a$, and $\zeta^c$ run all admissible values, is called the kernel of MGIs (principal group) of BVPs from the class under study. The respective Lie algebra
\[
A^{\mathrm{ker}} = \bigcap_{\theta, \vartheta^a, \zeta^c} A^{\mathrm{BVP}}_{\mathrm{max}}\left(F, \theta, \vartheta^a, \zeta^c \right).
\]
is called the algebra of kernel of MGIs.
\end{definition}

The group classification problem for the class under study is to find the kernel $G^{\mathrm{ker}}$ of MGIs of BVPs from the class (\ref{2.11})--(\ref{2.13}) and to describe all such cases of the arbitrary elements $\theta$, $\vartheta^a$, and $\zeta^c$ admitting nonequivalent expansions of the group $G^{\mathrm{ker}}$. We propose the following algorithm of the group classification for the class of BVPs (\ref{1})--(\ref{3}):
\begin{itemize}
\item[(I)] to construct the equivalence group $E_{\mathrm{eq}}$ of the class of partial differential equations systems~(\ref{2.11});
\item[(II)] to find the equivalence group $E_{\mathrm{eq}}^{\mathrm{BVP}}$ of the class of BVPs (\ref{2.11})--(\ref{2.13}) by extending the space of $E_{\mathrm{eq}}$--action on the prolonged  space, where all arbitrary elements $\vartheta^a$ and $\zeta^c$ arising in the boundary conditions (\ref{2.12}) and (\ref{2.13}) are treated as new dependent variables;
\item[(III)] to perform the group classification of the class of systems (\ref{2.11}) up to local transformations from the group $E_{\mathrm{eq}}^{\mathrm{BVP}}$;
\item[(IV)] to find the kernel $G^{\mathrm{ker}}$ of MGIs of BVPs from the class (\ref{2.11})--(\ref{2.13});
\item[(V)] to describe all possible $E_{\mathrm{eq}}^{\mathrm{BVP}}$--inequivalent BVPs of the form (\ref{2.11})--(\ref{2.13}) admitting MGIs of higher dimensionality than $G^{\mathrm{ker}}$.
\end{itemize}

The algorithm proposed leads to a list of nonequivalent (up to the local transformations from the equivalence group $E_{\mathrm{eq}}^{\mathrm{BVP}}$) BVPs of the form (\ref{2.11})--(\ref{2.13}) with the relevant MGIs (or MAIs). The problem of group classification is considered to solve in full if it was proved that:

\begin{itemize}
\item[i)] the groups (algebras) constructed are the MGIs (MAIs, respectively) of the BVPs obtained;

\item[ii)] all BVPs from the list are nonequivalent with respect to the transformations from the group  $E_{\mathrm{eq}}^{\mathrm{BVP}}$.
\end{itemize}

It should be noted that the class (\ref{0.1})--(\ref{0.4}) is parameterized by two arbitrary functions and a constant, so, to investigate symmetry properties of its, we will exploit the algorithm proposed.


\section{\bf Lie's symmetries of the class of BVPs (\ref{0.1})--(\ref{0.4})}

In this section we will perform the complete group classification of the class of BVPs (\ref{0.1})--(\ref{0.4}). In conformity with the algorithm proposed, at the first step we should seek for the equivalence group $E_{\mathrm{eq}}$ of the class of nonlinear heat equations (\ref{0.1}). However, this group is well-known \cite{Ovs2}
\begin{equation}\label{3.1}
\bar{t} = e_1 t + t_0, \ \ \bar{x} = e_2 x + x_0, \ \ \bar{u} = e_3 u + u_0, \ \ \bar{d} = \frac{e_2^2}{e_1} \, d,
\end{equation}
where $e_1, e_2, e_3, t_0, x_0, u_0 \in \mathbb{R}, \ e_1e_2e_3 \neq 0$.

\begin{lema}
The class of BVPs (\ref{0.1})--(\ref{0.4}) admits the equivalence group of point transformations~$E_{\mathrm{eq}}^{\mathrm{BVP}}$
\begin{equation}\label{3.3}
\tilde{t} = e_1 t + t_0, \ \ \tilde{x} = e_2 x, \ \ \tilde{u} = e_3 u + u_0,
\end{equation}
\begin{equation}\label{3.4}
\tilde{d} = \frac{e_2^2}{e_1} \, d, \ \tilde{q} = \frac{e_2 e_3}{e_1} \, q, \ \tilde{u}_{\infty} = e_3 u_{\infty} + u_0,
\end{equation}
where $e_1, e_2, e_3, t_0, u_0 \in \mathbb{R} \ (e_2 > 0, \ e_1 e_3 \neq 0)$.
\end{lema}
\textbf{Proof.} According to Definition~\ref{def4}, the group of equivalent transformations $E_{\mathrm{eq}}^{\mathrm{BVP}}$ is generated by all such point transformations of the independent and dependent variables mapping the class of BVPs (\ref{0.1})--(\ref{0.4}) into itself, i.e.
\begin{eqnarray}
& & \frac{\partial \tilde{u}}{\partial \tilde{t}} = \frac{\partial}{\partial \tilde{x}}\left(\tilde{d}(\tilde{u})\frac{\partial \tilde{u}}{\partial \tilde{x}}\right), \label{3.5} \\
& & \quad \tilde{x} = 0: \tilde{d}(\tilde{u})\frac{\partial \tilde{u}}{\partial \tilde{x}} = \tilde{q}({\tilde{t}}), \label{3.7} \\
& & \quad \tilde{x} = + \infty: \tilde{u} = \tilde{u}_{\infty}. \label{3.8}
\end{eqnarray}
Substituting transformations (\ref{3.1}) into the class of boundary conditions (\ref{3.7}) one obtains
\begin{equation}\label{3.6}
x = - \frac{x_0}{e_2}: \ d(u) \frac{\partial u}{\partial x} = \frac{e_1}{e_2 e_3} \, \tilde{q}.
\end{equation}
Taking into account (\ref{0.3}), from (\ref{3.6}) we immediately receive
\[ x_0 = 0, \ \ \tilde{q} = \frac{e_2 e_3}{e_1} \ q. \]
If in a quite similar way we consider the class of boundary conditions (\ref{3.8}), one can arrive at transformations (\ref{3.3}) and (\ref{3.4}).

The proof is now complete. \qquad \qquad \qquad \qquad \qquad \qquad \qquad \qquad \qquad \qquad \qquad \qquad \qquad $\blacksquare$

\begin{theorem}\label{T1}
All possible MAIs (up to the equivalent transformations from the group $E_{\mathrm{eq}}^{\mathrm{BVP}}$) of equation (\ref{0.1}) for any fixed strictly positive function $d(u) \neq \mbox{const}$ are presented in Table 1. Any other equation of the form (\ref{0.1}) is reduced by an equivalence transformation from the group $E_{\mathrm{eq}}^{\mathrm{BVP}}$ to one of those given in Table 1.
\end{theorem}

\begin{table}
\rightline{Table 1} \centerline{\bf The result of group classification of the class of nonlinear heat equations (\ref{0.1})}
\centerline{\bf up to the equivalent transformations from the group $E_{\mathrm{eq}}^{\mathrm{BVP}}$}
\medskip
{\renewcommand{\arraystretch}{1.5}
\begin{center}
\begin{tabular}{|c|l|l|}
  \hline
  Case & Eqs. (\ref{0.1}) & Basic operators of MAI \\
  \hline
  1.& $u_{t} = (d(u) u_{x})_{x}$ & $A^{\mathrm{ker}} = \langle \partial_{t}, \partial_{x}, 2t \partial_{t} + x \partial_{x} \rangle$  \\
  2. & $u_{t} = (e^u u_{x})_{x}$ & $\langle A^{\mathrm{ker}}, x \partial_{x} + 2 \partial_{u} \rangle$  \\
  3. & $u_{t} = (u^k u_{x})_{x}, k\neq 0, -\frac {4}{3}$ & $\langle A^{\mathrm{ker}}, k x \partial_{x} + 2 u \partial_{u} \rangle$ \\
  4. & $u_{t} = (u^{-\frac {4}{3}} u_{x})_{x}$ & $\langle A^{\mathrm{ker}}, -\frac {4}{3} x \partial_{x} + 2 u \partial_{u}, x^2 \partial_{x} - 3xu\partial_{u} \rangle$
\\
  \hline
\end{tabular}
\end{center}}
\end{table}

\noindent \textbf{Proof.} It follows directly from the result of group classification of the nonlinear heat equation carried out up to local transformations from the group $E_{\mathrm{eq}}$ \cite{Ovs2}.

The proof is now complete. \qquad \qquad \qquad \qquad \qquad \qquad \qquad \qquad \qquad \qquad \qquad \qquad \qquad $\blacksquare$

\begin{theorem}\label{T2}
All possible MGIs (up to equivalent transformations from the group $E_{\mathrm{eq}}^{\mathrm{BVP}}$) of the nonlinear BVP (\ref{0.1})--(\ref{0.4}) for any fixed set of arbitrary elements $(d(u), q(t), u_{\infty})$, where $d(u) \neq \mbox{const}$ are presented in Table 2. Any other BVP of the form (\ref{0.1})--(\ref{0.4}) is reduced by an equivalence transformation from the group $E_{\mathrm{eq}}^{\mathrm{BVP}}$ to one of those with the arbitrary elements given in Table~2.
\end{theorem}

\begin{remark}\label{rem1}
In Table 2, we used the following designations for the one-parameter Lie groups
\[
\begin{split}
& T_d: \ t^{\ast} = t e^{2 \varepsilon_d}, \ \ x^{\ast} = x e^{\varepsilon_d}, \ \ u^{\ast} = u;\\
& T_t: \ t^{\ast} = t + \varepsilon_t, \ \  x^{\ast} = x, \ \ u^{\ast} = u;\\
& T_k: \ t^{\ast} = t, \ \ x^{\ast} = x e^{k \varepsilon_k}, \ \ u^{\ast} = u e^{2 \varepsilon_k};\\
& T_{kp}: \ t^{\ast} = t e^{(k+2) \varepsilon_{kp}}, \ \ x^{\ast} = x e^{\left(k(p+1)+1\right) \varepsilon_{kp}}, \ \ u^{\ast} = u e^{(2p+1) \varepsilon_{kp}};\\
& T_{ke}: \ t^{\ast} = t + (k+2) \varepsilon_{ke}, \ \  x^{\ast} = x e^{k \varepsilon_{ke}}, \ \ u^{\ast} = u e^{2 \varepsilon_{ke}},
\end{split}
\]
where $\varepsilon$ with indexes are the relevant group parameters.
\end{remark}

\begin{remark}\label{rem2}
Up to local transformations from the group $E_{\mathrm{eq}}^{\mathrm{BVP}}$, the parameter $q_0 \neq 0$ may be set equal to $\pm 1$.
\end{remark}

\begin{table}
\rightline{Table 2} \centerline{\bf The result of group classification of the class of BVPs (\ref{0.1})--(\ref{0.4})}
\centerline{\bf up to the equivalent transformations from the group $E_{\mathrm{eq}}^{\mathrm{BVP}}$}
\medskip
{\renewcommand{\arraystretch}{1.5}
\begin{center}
\begin{tabular}{|c|c|c|c|l|l|}
  \hline
  Case & $d(u)$ & $q(t)$ & $u_{\infty}$ & MGI & Additional constraints \\
  \hline\hline
  1. & $\forall$ & $q_0 t^{-\frac{1}{2}}$ & $\forall$ & $T_d$  & \\
  2. & $\forall$ & $q_0$ & $\forall$ & $T_t$  & \\
  3. & $\forall$ & 0 & $\forall$ & $T_t, \ T_d$  & \\
  4. & $u^k$ & $q_0 t^p$ & 0 & $T_{kp}$  & $k \neq -2, \ p \neq - \frac{1}{2}$; \ $p \neq 0$ \\
  5. & $u^k$ & $q_0 e^t$ & 0 & $T_{ke}$ & $k \neq -2$\\
  6. & $u^k$ & $q_0$ & 0 & $T_t, \ T_{kp}$ & $p = 0$\\
  7. & $u^k$ & 0 & 0 & $T_t, \ T_d, \ T_k$ & \\
  8. & $u^{-2}$ & $\forall$ & 0 & $T_k$ & $k = -2$\\
  9. & $u^{-2}$ & $q_0 t^{- \frac{1}{2}}$ & 0 & $T_d, \ T_k$ & $k = -2$\\
  \hline
\end{tabular}
\end{center}}
\end{table}

\noindent \textbf{Proof.} Let us consider the case of the arbitrary function $d(u) \neq \mbox{const}$. In accordance with Theorem \ref{T1}, the MGI of the governing equation (\ref{0.1}) is the three-parameter Lie group generated by the one-parameter groups $T_t, T_d$, and the group
\[
T_x: \ t^{\ast} = t, \ \ x^{\ast} = x + \varepsilon_x, \ \ u^{\ast} = u.
\]
First, we will study the invariance of the BVPs from the class in question w.r.t. the Lie group $T_d$. According to item 2) of Definition \ref{def1}, the boundary conditions (\ref{0.3}) is invariant under $T_d$ if the conditions
\begin{equation}\label{16}
x^{\ast}\vert_{\cal M} = 0,  \ \ \left. d(u^{\ast})\frac{\partial u^{\ast}}{\partial x^{\ast}} - q(t^{\ast})\right \vert_{\cal M} = 0
\end{equation}
are satisfied on the manifold $\cal M$ = $ \left \{x = 0, \ d(u)\frac{\partial u}{\partial x} = q(t)\right \}$. The first
equation from (\ref{16}) is identically fulfilled, while the second one leads to the requirement
\[
e^{\varepsilon_d} q(e^{2 \varepsilon_d} t) = q(t),
\]
from which we immediately obtain $q(t) = q_0 t^{-\frac{1}{2}}$, where $q_0 \neq 0$ is an arbitrary real number, or $q(t) \equiv 0$. The invariance of the boundary condition (\ref{0.4}) w.r.t. the Lie group under study is evident. Thus, one can conclude that BVP (\ref{0.1})--(\ref{0.4}) with the arbitrary function d(u) is invariant w.r.t. $T_d$ if and only if $q(t) = q_0 t^{-\frac{1}{2}}$ (case 1 in Table 2), or $q(t) \equiv 0$.

It is easy to obtain that the BVP in question is not invariant w.r.t. the Lie group $T_x$, since the boundary curve $x =
0$ is not invariant under the point transformations from $T_x$. At the same time, the invariance of BVP (\ref{0.1})--(\ref{0.4}) w.r.t. $T_t$ leads to the requirement $q(t) = \mbox{const}$ (case~2 in Table 2).

In order to complete examination of the case of the arbitrary function $d(u)$, let us study the invariance of the BVPs from class (\ref{0.1})--(\ref{0.4}) w.r.t. the one-parameter Lie groups corresponding to the linear combination of the basic generators of the algebra $A^{\mathrm{ker}}$ of equation (\ref{0.1})
\[
X = (\lambda_1 + 2 \lambda_3 t) \partial_t + (\lambda_2 + \lambda_3 x) \partial_x,
\]
where $\lambda_i, \ i = 1, \ldots, 3$ are arbitrary real parameters.

Let $\lambda_3 = 0$, then the Lie group $T_{xt}$
\[
T_{xt}: \ t^{\ast} = t + \lambda_1 \varepsilon_{xt}, \ \ x^{\ast} = x + \lambda_2 \varepsilon_{xt}, \ \ u^{\ast} = u.
\]
corresponds to the generator $X$. For the boundary curve $x = 0$ to be invariant w.r.t. $T_{xt}$, the condition $\lambda_2 = 0$ must hold. Therefore, the group $T_{xt}$ is reduced to $T_{t}$, which we completely considered above.

Now let $\lambda_3 \neq 0$. In this case, the operator $X$ generates the one-parameter Lie group
\[
T_{X}: \ t^{\ast} = t e^{2 \lambda_3 \varepsilon_{X}} + \frac{\lambda_1}{2 \lambda_3} \left(e^{2 \lambda_3 \varepsilon_{X}} - 1 \right), \ \ x^{\ast} = x e^{\lambda_3 \varepsilon_{X}} + \frac{\lambda_2}{\lambda_3} \left(e^{\lambda_3 \varepsilon_{X}} - 1 \right), \ \ u^{\ast} = u.
\]
The invariance of the boundary condition (\ref{0.4}) w.r.t. $T_{X}$ is evident, while the invariance of the boundary condition
(\ref{0.1}) leads to the restrictions $\lambda_2 = 0$ and $q(t) = q_0 \left(t + \frac{\lambda_1}{2 \lambda_3} \right)^{-
\frac{1}{2}}$. Up to local transformations from $E_{\mathrm{eq}}^{\mathrm{BVP}}$, one can set $t + \frac{\lambda_1}{2 \lambda_3} \rightarrow t$ and reduce the group $T_{X}$ to $T_{d}$, with the function $q(t)$ taking the form $q(t) = q_0 t^{-\frac{1}{2}}$.

Thus, any linear combination of the basic generators of the nonlinear heat equation (\ref{0.1}) with the arbitrary function $d(u)$ does not lead to any new one-parameter group of invariance.

Taking into account the invariance conditions of the BVPs from the class under study w.r.t. the groups $T_t$ and $T_d$, one obtains that BVP (\ref{0.1})--(\ref{0.4}) is invariant under the two-parameter Lie group $T_t \circ T_d$ if and only if $q(t) \equiv 0$ (case 3 in Table 2). It means that the case of the arbitrary function $d(u)$ is completely investigated.

Let us now consider case 2 from Table 1, i.e. when $d(u) = e^{u}$. Here the nonlinear heat equation (\ref{0.1}) is invariant w.r.t. the four-parameter Lie group generated by the one-parameter groups $T_t$, $T_x$, $T_d$, and the group
\begin{equation}\label{a1}
T_e: \ t^{\ast} = t, \ \ x^{\ast} = x e^{\varepsilon_e}, \ \ u^{\ast} = u + 2 \varepsilon_e.
\end{equation}
Obviously, the BVPs from class (\ref{0.1})--(\ref{0.4}) are not invariant w.r.t. the Lie group $T_e$. Indeed, for the boundary condition (\ref{0.4}) to be invariant under $T_e$, the following conditions must satisfy
\[
x^{\ast}\vert_{\cal N} = + \infty, \ \ u^{\ast}\vert_{\cal N} = u_{\infty},
\]
for an arbitrary value $\varepsilon_e \in \mathbb{R}$, where $\cal N$ = $ \{x = + \infty, \ u = u_{\infty} \}$ . Taking into account (\ref{a1}), from the second equality one immediately obtains $\varepsilon_e = 0$. Thus, the contradiction is obtained and we can conclude that class (\ref{0.1})--(\ref{0.4}) is not invariant w.r.t. $T_e$.

In a quite similar way one can show that the BVPs from the class in question is not invariant w.r.t. the one-parameter Lie groups corresponding to any linear combination of the generator $X_e = x \partial_x + 2\partial_u$ with the generators of the algebra $A^{\mathrm{ker}}$. Thereby, we conclude that the case $d(u) = e^{u}$ does not lead to any new group of invariance.

Let us consider the most interesting case 3 from Table 1, i.e. $d(u) = u^k, \ k \neq -\frac{4}{3}$. In this case, equation (\ref{0.1}) is invariant w.r.t. the four-parameter Lie group, generated by the one-parameter groups $T_t$, $T_x$, $T_d$, and the group
\[
T_k: \ t^{\ast} = t, \ \ x^{\ast} = x e^{k \varepsilon_k}, \ \ u^{\ast} = u e^{2 \varepsilon_k}.
\]
First, we will investigate the invariance of the BVPs from class (\ref{0.1})--(\ref{0.4}) w.r.t. the group $T_k$. According to item 2) of Definition \ref{def1}, the boundary condition
(\ref{0.3}) is invariant under $T_k$ if, on the manifold $\cal P$ = $ \{x = 0, \ u^k \frac{\partial u}{\partial x} - q(t) = 0\}$, the following equalities
\begin{equation}\label{dod1}
\left. x^{\ast} \right \vert_{\cal P} = 0, \ \left. (u^{\ast})^k \frac{\partial u^{\ast}}{\partial x^{\ast}} - q(t^{\ast}) \right \vert_{\cal P}= 0
\end{equation}
are true. From (\ref{dod1}) one obtains the relation
\[ q(t) e^{(k+2) \varepsilon_k} = q(t),
\]
which immediately leads to the conditions: $k = -2$, $q(t)$ is arbitrary (case 8 in Table 2) and $k \in \mathbb{R}$, $q(t) = 0$. The boundary condition (\ref{0.4}) is invariant w.r.t. $T_k$ if and only if $u_{\infty} = 0$.

Taking into account the invariance of BVPs (\ref{0.1})--(\ref{0.4}) w.r.t. the one-parameter groups $T_t, T_x$, and $T_k$,
one can draw a conclusion that it are invariant w.r.t. the two-parameter Lie group $T_d \circ T_k$ (case 9 in Table 2) and the
three-parameter Lie group $T_t \circ T_d \circ T_k$ (case 7 in Table 2) if and only if the following conditions
\[
d(u) = u^{-2}, \ \ q(t) = \delta t^{-\frac{1}{2}}, \ \ u_{\infty} = 0.
\]
and
\[
d(u) = u^{k}, \ \ q(t) = 0, \ \ u_{\infty} = 0.
\]
hold, respectively.

Let us now consider the invariance of the BVPs from the class under study w.r.t. the one-parameter Lie groups corresponding to any linear combination of the generator $k x \partial_x + 2 u \partial_u \ (k \neq 0, -4/3)$ with the generators from the algebra $A^{\mathrm{ker}}$
\[
Y = (\lambda_1 + 2 \lambda_3 t) \partial_t + (\lambda_2 + (\lambda_3 + k) x) \partial_x + 2 u \partial_u,
\]
where $\lambda_i, \ i = 1, \ldots, 3$ are arbitrary real parameters. The operator $Y$ generates depending from the values of the parameters $\lambda_i$ and $k$ such one-parameter Lie groups
\begin{itemize}
\item[1)] $\lambda_3 \neq 0, \ \lambda_3 + k \neq 0$
\[
T_{1}: \ t^{\ast} = t e^{2 \lambda_3 \varepsilon_{1}} + \frac{\lambda_1}{2 \lambda_3} \left(e^{2 \lambda_3 \varepsilon_{1}} - 1 \right), \ \ x^{\ast} = x e^{(\lambda_3 + k)  \varepsilon_{1}} + \frac{\lambda_2}{\lambda_3 + k} \left(e^{(\lambda_3 + k) \varepsilon_{1}} - 1 \right), \ \ u^{\ast} = u e^{2 \varepsilon_{1}};
\]
\item[2)] $\lambda_3 \neq 0, \ \lambda_3 + k = 0$
\[
T_{2}: \ t^{\ast} = t e^{-2 k \varepsilon_{2}} - \frac{\lambda_1}{2 k} \left(e^{-2 k \varepsilon_{2}} - 1 \right), \ \ x^{\ast} = x + \lambda_2 \varepsilon_{2}, \ \ u^{\ast} = u e^{2 \varepsilon_{2}};
\]
\item[3)] $\lambda_3 = 0$
\[
T_{3}: \ t^{\ast} = t + \lambda_1 \varepsilon_{3}, \ \  x^{\ast} = x e^{k \varepsilon_{3}} + \frac{\lambda_2}{k} \left(e^{k \varepsilon_{3}} - 1 \right), \ \ u^{\ast} = u e^{2 \varepsilon_{3}}.
\]
\end{itemize}

First, we will study the case of the group $T_1$. The invariance of the boundary conditions (\ref{0.3}) w.r.t. $T_1$ (see equalities (\ref{dod1})) gives
\begin{equation}\label{3.14}
\lambda_2 = 0, \ \ q(t) e^{(k + 2 - \lambda_3) \varepsilon_1} = q \left(t e^{2 \lambda_3 \varepsilon_1} + \frac{\lambda_1}{2 \lambda_3} \left(e^{2 \lambda_3 \varepsilon_{1}} - 1 \right) \right).
\end{equation}
The second equality of (\ref{3.14}) leads to the condition at the function $q(t)$
\begin{equation}\label{3.15}
q(t) = q_0 \left(t + \frac{\lambda_1}{2 \lambda_3} \right)^{\frac{k + 2 -\lambda_3}{2 \lambda_3}},
\end{equation}
where $q_0 \neq 0$ is an arbitrary real constant.
Making in (\ref{3.15}) the change of the variable $t + \frac{\lambda_1}{2 \lambda_3} \rightarrow t$, one obtains
\[
q(t) = q_0 t^{\frac{k + 2 -\lambda_3}{2 \lambda_3}},
\]
while the transformations from the group $T_1$ get such form
\[
T'_{1}: \ t^{\ast} = t e^{2 \lambda_3 \varepsilon_{1}}, \ \  x^{\ast} = x e^{(\lambda_3 + k) \varepsilon_{1}}, \ \ u^{\ast} = u e^{2 \varepsilon_{1}}.
\]
Let us now reduce the form of function $q(t)$ using the substitution $p = \frac{k +2 -\lambda_3}{2 \lambda_3}$, from which we can easy obtain $\lambda_3 = \frac{k + 2}{2 p + 1}$. Plugging this value of $\lambda_3$ into $T'_1$, one arrives at the group $T_{kp}$ (see Remark \ref{rem1}). The invariance of the boundary condition (\ref{0.4}) w.r.t. $T_{kp}$ gives directly $u_{\infty} = 0$.

Thus, we proved that the BVP from class (\ref{0.1})--(\ref{0.4}) is invariant w.r.t. the one-parameter Lie group $T_{kp}$ if and only if the arbitrary elements $d(u), q(t)$, and $u_{\infty}$ have such values
\[
d(u) = u^k (k \neq -2), \ \ q(t) = q_0 t^p (p \neq - \frac{1}{2}), \ \ u_{\infty} = 0.
\]
It is exactly case 4 in Table 2.

The case of the group $T_2$ is considered in a quite similar way, and as a result one obtains the conditions
\[
\lambda_2 = 0, \ \ q(t) = q_0 t^{-\frac{k+1}{k}}, \ \ u_{\infty} = 0;
\]
while the group $T_2$ reduces to the form
\[
T'_{2}: \ t^{\ast} = t e^{-2 k \varepsilon_{2}}, \ \  x^{\ast} = x, \ \ u^{\ast} = u e^{2 \varepsilon_{2}}.
\]
It is easy to see if we set $p = -\frac{k+1}{k}$ than the group $T'_2$ is a particular case of the group $T_{kp}$. Hence, the case of the group $T_2$ does not lead to any new group of invariance of the BVPs from the class under study.

Consider the case of the group $T_3$. The invariance of the boundary condition (\ref{0.4}) w.r.t. $T_3$ gives again the condition $u_{\infty} = 0$, while, from the invariance of the boundary conditions (\ref{0.3}), we arrive at the restrictions
\begin{equation}\label{3.16}
\lambda_2 = 0, \ \ q(t) e^{(k+2) \varepsilon_3} = q(t + \lambda_1 \varepsilon_3).
\end{equation}
It follows from the second equality in (\ref{3.16}) that
\[
q(t) = q_0 e^{\frac{k+2}{\lambda_1}t},
\]
where $q_0 \neq 0$ is an arbitrary real constant.
Making the change of the variable $\frac{k+2}{\lambda_1}t \rightarrow t$, one obtains that the group $T_3$ is reduced to the group $T_{ke}$. Hence, the BVP from class (\ref{0.1})--(\ref{0.4}) is invariant w.r.t. the group  $T_{ke}$ if and only if the arbitrary elements have such form
\[
d(u) = u^k (k \neq -2), \ \ q(t) = q_0 e^t, \ \ u_{\infty} = 0,
\]
Thus, case 5 in Table 2 is obtained.

Taking into account the invariance conditions of the BVP from class (\ref{0.1})--(\ref{0.4}) w.r.t. the one-parameter Lie groups  $T_t, T_d, T_k, T_{kp}$, and $T_{ke}$, we can conclude that such BVP is invariant w.r.t. the two-parameter Lie group $T_t \circ T_{kp} (p = 0)$ (case 6 in Table 2) if and only if
\[
d(u) = u^k, \ \ q(t) = q_0, \ \ u_{\infty} = 0.
\]
Thus, the case of the function $d(u) = u^k \ (k \neq 0, -4/3)$ is completely examined.

Let us now consider the case of the conformal power $k = -\frac{4}{3}$, i.e. $d(u) = u^{-\frac{4}{3}}$. According to Theorem \ref{T1}, in this case equation (\ref{0.1}) is invariant w.r.t. the five-parameter Lie group generated by the one-parameter groups $T_t, T_x, T_d, T_k (k = -\frac{4}{3})$, and the group
\[
T_c: \ t^{\ast} = t, \ \ x^{\ast}  = \frac{x}{1 - \varepsilon_c x}, \ \ u^{\ast} = (1 - \varepsilon_c x)^3 u.
\]
First, we stress that the one-parameter groups (with $k = - \frac{4}{3}$) listed in cases 4--7 of Table 2 are the groups of invariance of the BVPs in question (with $k = - \frac{4}{3}$) under the same restrictions on the function $q(t)$ and constant $u_{\infty}$. Now, our purpose is to show that the group $T_c$ does not lead to any case of invariance of BVPs from class (\ref{0.1})--(\ref{0.4}). Indeed, for the boundary condition (\ref{3.3}) to be invariant w.r.t. $T_c$, the following conditions must hold
\[
x^{\ast}\vert_{\cal N} = + \infty, \ \ u^{\ast}\vert_{\cal N} = u_{\infty},
\]
for arbitrary value of $\varepsilon_c \in \mathbb{R}$, where $\cal N$ = $ \{x = + \infty, \ u = u_{\infty} \}$. However
\[
\lim_{x \rightarrow + \infty} x^{\ast} = \lim_{x \rightarrow + \infty} \frac{x}{1 - \varepsilon_c x} = - \frac{1}{\varepsilon_c}.
\]
Therefore, the condition $x^{\ast}\vert_{\cal N} = + \infty$ is satisfied if and only if the group parameter $\varepsilon_c = 0$. Thus, the contradiction is obtained, and we can draw a conclusion that the BVPs from class (\ref{0.1})--(\ref{0.4}) are not invariant w.r.t. $T_c$.

Let us now examine the invariance the BVPs from the class under study w.r.t. the one-parameter Lie groups corresponding to any linear combination of the conformal operator $x^2 \partial_x - 3 x u \partial_u$ with the operators of the algebra $\langle A^{\mathrm{ker}}, -\frac {4}{3} x \partial_{x} + 2 u \partial_{u} \rangle$
\[
Z= (\lambda_1 + 2 \lambda_3 t) \partial_t + \left(\lambda_2 + \left(\lambda_3 - \frac{4}{3} \lambda_4\right) x + x^2\right) \partial_x + (2 \lambda_4 - 3 x) u \partial_u,
\]
where $\lambda_i, \ i = 1, \ldots, 4$ are arbitrary real constants.
First, we note that the invariance of the boundary curve $x = 0$ w.r.t. the generator $Z$ immediately yields $\lambda_2 = 0$. Let now the following conditions $\lambda_3 \neq 0, \lambda_4 \neq 0, \lambda_3 -\frac{4}{3} \lambda_4 \neq 0$ hold. In this case, the transformation of the variable $x$ from the corresponding group has such form
\begin{equation}\label{3.17}
x^{\ast} = \frac{x \left(\lambda_3 - \frac{4}{3} \lambda_4\right) e^{\left(\lambda_3 - \frac{4}{3} \lambda_4\right) \varepsilon_Z}}{x \left(1 - e^{\left(\lambda_3 - \frac{4}{3} \lambda_4\right) \varepsilon_Z} \right) + \lambda_3 - \frac{4}{3} \lambda_4}.
\end{equation}
Taking into account (\ref{3.17}), the equality $x^{\ast}\vert_{\cal N} = + \infty$ gives
\[
\lim_{x \rightarrow + \infty} x^{\ast} = \lim_{x \rightarrow + \infty} \frac{x \left(\lambda_3 - \frac{4}{3} \lambda_4\right) e^{\left(\lambda_3 - \frac{4}{3} \lambda_4\right) \varepsilon_Z}}{x \left(1 - e^{\left(\lambda_3 - \frac{4}{3} \lambda_4\right) \varepsilon_Z} \right) + \lambda_3 - \frac{4}{3} \lambda_4} = \frac{\left(\lambda_3 - \frac{4}{3} \lambda_4\right) e^{\left(\lambda_3 - \frac{4}{3} \lambda_4\right) \varepsilon_Z}}{1 - e^{\left(\lambda_3 - \frac{4}{3} \lambda_4\right) \varepsilon_Z}}.
\]
The cases when at least one of the parameters $\lambda_3$, or $\lambda_4$ are not zero can be examined in a quite similar way, whereas the cases $\lambda_3 = \lambda_4 = 0$, or $\lambda_3 -\frac{4}{3} \lambda_4 = 0$ lead to the group $T_c$ considered above.

Thus, we showed that any BVP from class (\ref{0.1})--(\ref{0.4}) is not invariant w.r.t. each one-parameter Lie group corresponding to any linear combination of the conformal operator $x^2 \partial_x - 3 x u \partial_u$ with the operators from the MAI of equation (\ref{0.1}) with $d(u) = u^{-\frac{4}{3}}$.


The proof is now complete. \qquad \qquad \qquad \qquad \qquad \qquad \qquad \qquad \qquad \qquad \qquad \qquad \qquad $\blacksquare$

\section{\bf Symmetry reduction and invariant solutions of a BVP from class (\ref{0.1})--(\ref{0.4})}

In this section we will use the results of the group classification of the class of BVPs (\ref{0.1})--(\ref{0.4})
obtained in the previous section for reduction one BVP from the class and construction an exact invariant solution for its. Let us restrict ourselves to the case 6 from Table 2. Another words, we consider a BVP of the form
\begin{eqnarray}
& & \frac{\partial u}{\partial t} = \frac{\partial}{\partial x}\left(u^k \frac{\partial u}{\partial x}\right), \ \ t > 0, \ x > 0, \label{3.18} \\
& & \quad x = 0: u^k \frac{\partial u}{\partial x} = q_0, \ \  t > 0, \label{3.19} \\
& & \quad x = + \infty: u = 0, \ \ t > 0. \label{3.20}
\end{eqnarray}
According to Theorem 2, BVP (\ref{3.18})--(\ref{3.20}) admits the two-parameter Lie group $G^6_2$ induced by the following
one-parameter groups
\[T_{k0}: \ t^{\ast} = t e^{(k + 2) \varepsilon_{k0}}, \ \ x^{\ast} = x e^{(k + 1) \varepsilon_{k0}}, \ \ u^{\ast} = u e^{ \varepsilon_{k0}},\]
\[T_t: \ t^{\ast} = t + \varepsilon_t, \ \ x^{\ast} = x, \ \ u^{\ast} = u. \]
It is easy to obtain that the two-dimensional Lie algebra
\[
X_{k0} = (k + 2) t \partial_t + (k + 1) x \partial_x  +  u \partial_u, \ \ X_t = \partial_t
\]
corresponds to the group $G^6_2$.

Consider the reduction of the BVP in question by the generator $X_{k0}$. Using the standard procedure, one can easy obtain the ansatz
\begin{equation}\label{4.2}
u = t^{\frac{1}{k + 2}} F(\omega), \ \ \omega = x t^{-\frac{k + 1}{k + 2}}.
\end{equation}
Substituting (\ref{4.2}) in BVP (\ref{3.18})--(\ref{3.20}), and making relevant calculations, we arrive at the reduced~BVP
\begin{eqnarray}
& & (F^k F_{\omega})_{\omega} + \frac{k + 1}{k+2} \ \omega F_{\omega} - \frac{1}{k+2} \ F = 0, \label{4.3} \\
& & \quad \omega = 0: \ F^k F_{\omega} = 1, \label{4.4} \\
& & \quad \omega = + \infty: \ F = 0.\label{4.5}
\end{eqnarray}

Now one sees that the problem of construction of exact solutions of BVP (\ref{3.18})--(\ref{3.20}) is reduced to solving the non-linear ordinary differential equation (\ref{4.3}). If the parameter $k$ is arbitrary, than a general solution of equation (\ref{4.3}) is unknown, but several particular solutions of ones, for some values of the parameter $k$, are presented in the handbook \cite{Pol} (see, also, \cite{Hill}).

Here we will consider only one instance when $k = -\frac{3}{2}$. In this case, equation (\ref{4.3}) has the general solution written in the parametric form
\begin{equation}\label{4.10}
\omega = \frac{C_1^{3}}{2}  E^{-1} \sqrt{\frac{\tau + 1}{\tau}}, \ \ F = 4 C_1^{-4} E^2, \ \ \tau \in \mathbb{R}^{+},
\end{equation}
where $E = 1 + \sqrt{\frac{\tau + 1}{\tau}} \left(C_2 - \ln \left(\sqrt{\tau} + \sqrt{\tau + 1}\right) \right)$.
\\ Substituting (\ref{4.10}) into the boundary conditions (\ref{4.4}) and (\ref{4.5}), and making relevant calculations, we arrive at the solution of BVP (\ref{4.3})--(\ref{4.5}) with $k = -\frac{3}{2}$
\begin{equation}\label{4.11}
\begin{split}
& \omega = \frac{- 4 \sqrt{\tau + 1}}{q_0^3 \left(\sqrt{\tau + 1} \ln \left(\sqrt{\tau} + \sqrt{\tau + 1}\right) - \sqrt{\tau}\right)},\\
& F = \frac{q_0^4}{4} \left(1 - \sqrt{\frac{\tau + 1}{\tau}}  \ln \left(\sqrt{\tau} + \sqrt{\tau + 1}\right)  \right)^2, \  \tau \in \mathbb{R}^{+}.
\end{split}
\end{equation}

Now, in accordance with (\ref{4.2}), formulae (\ref{4.11}) lead to the solution in implicit form of BVP (\ref{3.18})--(\ref{3.20})
\[
u = \frac{q_0^4 t^2}{4} \left(1 - \sqrt{\frac{\tau + 1}{\tau}} \ln \left(\sqrt{\tau} + \sqrt{\tau + 1}\right) \right)^2,
\]
where $ \frac{- 4 \sqrt{\tau + 1}}{q_0^3 \left(\sqrt{\tau + 1} \ln \left(\sqrt{\tau} + \sqrt{\tau + 1}\right) - \sqrt{\tau}\right)} = x t$, $\tau \in \mathbb{R}^{+}$.

\begin{figure}
\begin{center}
\vspace{1cm}
\includegraphics[width=8cm]{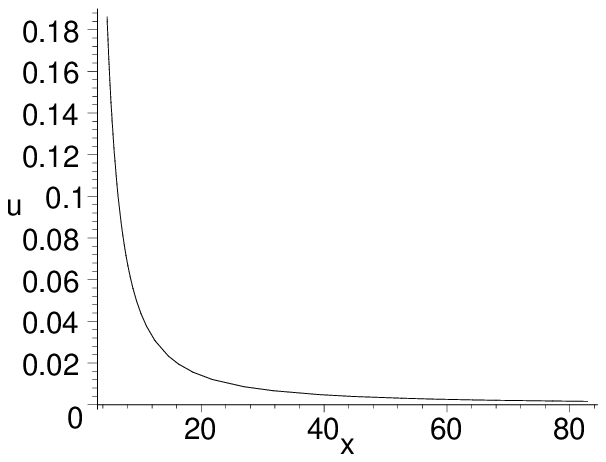}
\includegraphics[width=8cm]{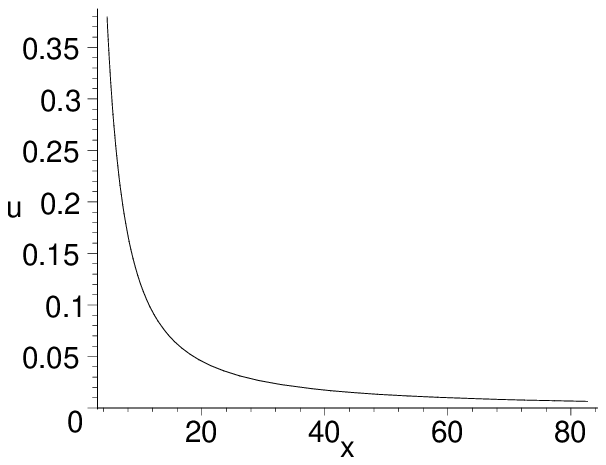}
\caption{Exact solutions of BVP (\ref{3.18})--(\ref{3.20}) with $k = -\frac{3}{2}$, $q_0 = - 1$ at the moments of time: 1)~$t = 0.5$ and 2)~$t = 5$.}
\end{center}
\end{figure}

\section{\bf Conclusions}

In this paper we proposed a new definition of invariance in Lie's sense of a BVP for systems of evolution differential equations. The definition formulated is essential extension of the known definitions of BVP invariance available in the literature. It is applicable to a wide class of BVPs and can be easily generalized to BVPs with hyperbolic and elliptic basic equations. The problem of group classification in a class of BVPs parameterized by arbitrary elements is formulated, and an algorithm for its solution is also proposed.

The main result of the paper is the complete group classification (carried out in Section 3) of the class of BVPs (\ref{0.1})--(\ref{0.4}) with (1+1)--dimensional governing nonlinear heat equation. It should be stressed that the nonlinear heat equation (\ref{0.1}) with the thermal conductivity coefficient depending on temperature as a power law, i.e. $d(u) = u^k$, has the MAI for the conformal power $k = - \frac{4}{3}$. However, BVP (\ref{0.1})--(\ref{0.4}) has the particular power $k = -2$ (see cases 8--9 in Table 2), which is not a special one for equation (\ref{0.1}). This fact emphasizes nontrivial character of the problem of investigation of symmetry properties for partial differential equations with both boundary and initial conditions.

We used the results of the group classification obtained in Section 3 to reduce a BVP from the class under study to a BVP of lower dimensionality. An example of its exact solution is also constructed.

\end{document}